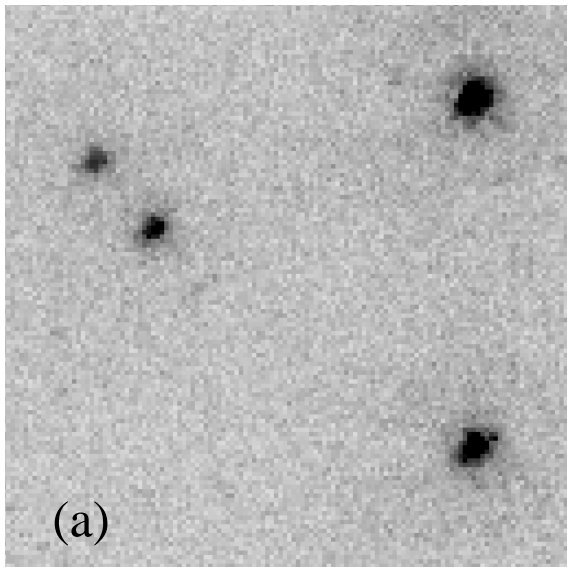

(a)

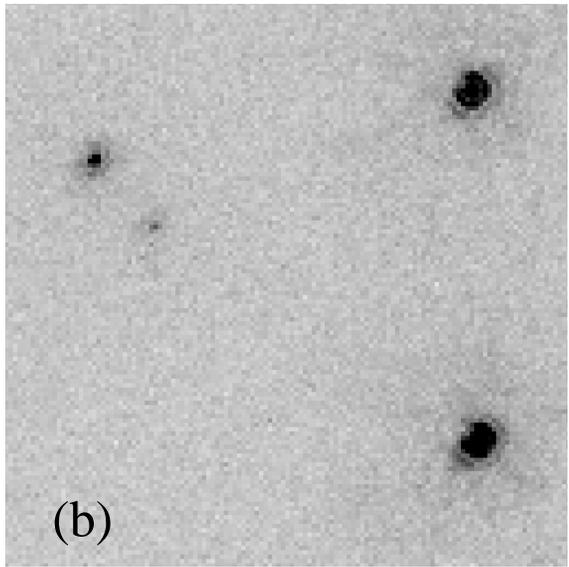

(b)

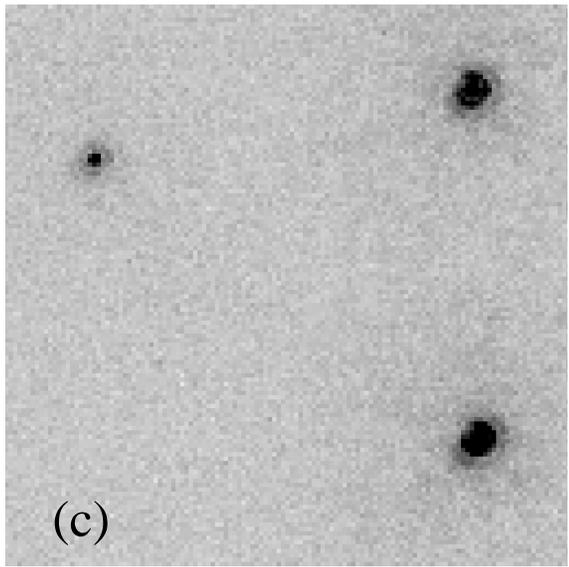

(c)

Fruchter, Bookbinder and Bailyn (1995)

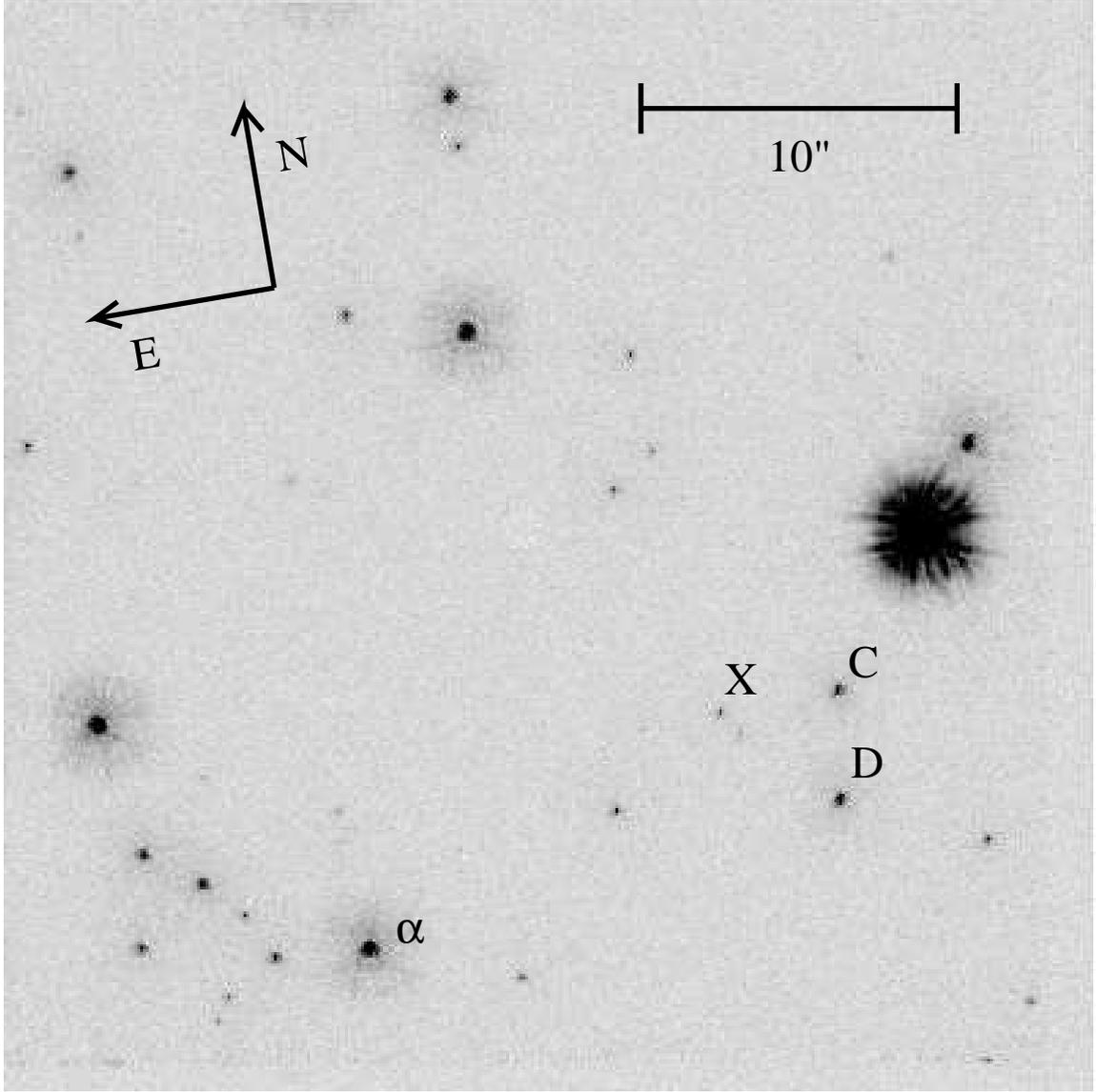

# HST Images of the Eclipsing Pulsar B1957+20

Andrew S. Fruchter
Hubble Fellow
Department of Astronomy
University of California
Berkeley, CA 94720
and
Space Telescope Science Institute
3700 San Martin Drive, Baltimore, MD 21218
fruchter@stsci.edu

Jay Bookbinder
Smithsonian Astrophysical Observatory
Cambridge, MA 02138
bookbind@cfa227.harvard.edu

Charles D. Bailyn
Department of Astronomy
Yale University
New Haven, CT 06511
bailyn@astro.yale.edu

## ABSTRACT

We have obtained images of the eclipsing pulsar binary PSR B1957+20 using the Planetary Camera of the Hubble Space Telescope. The high spatial resolution of this instrument has allowed us to separate the pulsar system from a nearby background star which has confounded ground-based observations of this system near optical minimum. Our images limit the temperature of the backside of the companion to $T \lesssim 2800$ K, about a factor of two less than the average temperature of the side of the companion facing the pulsar, and provide a marginal detection of the companion at optical minimum. The magnitude of this detection is consistent with previous work which suggests that the companion nearly fills its Roche lobe and is supported through tidal dissipation.

*Subject headings:* pulsars: eclipsing, millisecond, binary, PSR B1957+20 – stars: low-mass

– 2 –## 1. Introduction

The PSR B1957+20 binary (Fruchter, Stinebring & Taylor 1988) is an extraordinarily rich astrophysical system. One of the two fastest known pulsars, PSR B1957+20 is eclipsed at meter wavelengths for approximately ten percent of its 9.2 hour orbit by material expelled from its $\sim 0.025$ $M_\odot$ companion (Fruchter, Stinebring & Taylor 1988; Fruchter *et al.* 1990). Shortly before eclipse and for as long as one half-hour after eclipse, the radio pulses are measurably delayed, permitting an estimate of the electron density in the eclipsing wind (Fruchter, Stinebring & Taylor 1988; Fruchter *et al.* 1990; Ryba & Taylor 1991). Broad band optical observations show a dramatically variable system which is brightest when the side of the companion heated by the pulsar wind faces earth, and which darkens by several magnitudes when the cool, backside comes into view (Fruchter *et al.* 1988; Djorgovski & Evans 1988; van Paradijs *et al.* 1988; Callanan, Charles & van Paradijs 1989). The power of the pulsar's radiation is further displayed by a "cometary" H$\alpha$ nebula nearly a parsec across, formed by the interaction of the pulsar wind and the interstellar medium (Kulkarni & Hester 1988).

At the time of its discovery, it was thought that PSR B1957+20 might provide the missing link between low mass x-ray binaries (the presumed progenitors of all millisecond pulsars) and solitary millisecond pulsars; millisecond pulsars might evaporate their companions (Fruchter, Stinebring & Taylor 1988; Ruderman, Shaham & Tavani 1989; Kluźniak *et al.* 1988). However, VLA continuum images reveal that the eclipsing wind is translucent to 20 cm continuum radiation (Fruchter & Goss 1992). The wind is therefore far too tenuous to account for significant mass loss, unless it is overwhelmingly neutral (highly unlikely given the large systemic velocities and the power of the pulsar's high energy radiation) or unless the majority of mass loss is hidden from our view. The latter could perhaps occur were the companion to nearly fill its Roche lobe. Material could then leave the system at low velocity, remain in the orbital plane, and not necessarily cross our line of sight (Fruchter & Goss 1992; Banit & Shaham 1992).

A companion which nearly fills its Roche lobe also has the advantage of explaining the observed orbital period variability of PSR B1957+20 (Ryba & Taylor 1991; Arzoumanian, Fruchter & Taylor 1994). Such a companion would be convective and its rotation would be tidally locked to its orbit. As a result, it would have a Rossby number less than one and would be expected to display magnetic activity and orbital period variablity (Simon 1990) such as that seen in RS CVn binaries and Algol systems containing a convective member (Söderhjelm 1980; Hall 1989). Furthermore, tidal forces caused by mass loss could keep the rotation of the companion from becoming synchronized with the orbital period (Banit & Shaham 1992; Applegate & Shaham 1994). The resulting dissipation could provide the

– 3 –

energy to keep the companion bloated.

However, one would naively expect the 0.02 $M_\odot$ star to be a degenerate hydrogen dwarf, and calculations show that it should then have a radius about one-half that of its Roche lobe (Phinney *et al.* 1988) – and thus a gravitational potential well too deep to allow a low velocity wind. And, indeed, for quite a while it appeared that observations agreed with the prediction of a small companion. Optical images provide the companion's color temperature and apparent magnitude. Given an estimated distance, one can then derive the radius of the companion. The pulsar's dispersion measure was originally thought to imply a distance of 0.8 kpc and a companion radius of 0.12 $R_\odot$ – exactly one-half the Roche lobe radius (Fruchter *et al.* 1988; Djorgovski & Evans 1988). But a recent re-evaluation of the local electron density (Taylor & Cordes 1993) doubles the estimated distance to the pulsar and with it the estimated size of the companion (Fruchter & Goss 1992). The increased distance and companion size estimates are further supported by the results of long-slit spectroscopic observations of the H$\alpha$ nebula (Aldcroft, Romani & Cordes 1992) as well as detailed study of the pulsar's optical light curve (Callanan, van Paridijs & Regelink 1994); however, interpretation of the spectroscopic results is model dependent, and the accuracy of light-curve studies has so far been limited by the inability to measure the magnitude and colors of the companion at minimum due to the presence of a $\sim$ 21st magnitude star only 0$''$.8 on the sky away from PSR B1957+20. It was to remove the latter impediment that the work described here was undertaken.

## 2. Observations and Reduction

PSR B1957+20 was observed on three separate occasions using the Plantary Camera of the Hubble Space Telesecope (Mackenty *et al.* 1992). During the first of these observations we obtained a single 600s exposure of the system at maximum (orbital phase in the pulsar timing system of $\phi = 0.75$) using the F569W filter, which corresponds roughly to the Johnson $V$ band. Later, two sets of two 500 s exposures were planned to be taken at orbital phase 0.25, using the F791W filter, which covers a region of the spectrum comparable to the Cousins I band. Unfortunately, an error in the scheduling of the telescope caused the first of these sets of images to be taken at orbital phase 0.44. The second set was however correctly exposed near orbital phase 0.25, corresponding to pulsar eclipse and optical minimum. The times and phases of the images can be found in Table 1. All of these observations were taken before the refurbishment of the Space Telescope's optics.

Figure 1 displays a section of the images surrounding the pulsar after cosmic ray removal. The pixels at the position of PSR B1957+20 were not struck by cosmic rays in any of the exposures, so the cosmic ray removal techniques should not affect the science discussed here. In Figure 1a the system is seen in the "V" band near orbital phase $\phi = 0.75$, that is optical maximum. Figures 1b and 1c show the "I" band images centered at orbital phases $\phi = 0.44$ and $\phi = 0.27$, respectively. Although the filter used to obtain Figure 1a differs from that used for the other two images, the star to the north-east of the pulsar has nearly identical colors to that of the PSR B1957+20 at maximum, and therefore provides a fortuitous visual reference. It is this star, lying about 0".8 from the pulsar system on the sky, which has confounded ground-based observations of this system near optical minimum. The combined light of this object and the optical counterpart of PSR B1957+20 correspond to Star X on the finding chart of Kulkarni, Djogovski and Fruchter (1988). Throughout the remainder of this *Letter*, however, we will use the term Star X to refer to the background star alone.

Although these images provide dramatic evidence of the variability of PSR B1957's companion, they cannot be used to accurately limit the magnitude of the companion at miminimum without further processing, because the pulsar lies in the wings of the point spread function (PSF) of Star X. At the position of the pulsar, Star X contributes about one count (corresponding to about 7 photons) per pixel in each of the 500 s images. This is slightly more than the dark sky. In order to remove this background we must subtract the light from Star X using a PSF determined from a nearby bright star. However, the choice of a star as a PSF template involves a compromise, for the nearest unsaturated stars, C and D (see Figure 2), are, respectively, only three and four times brighter than Star X. As a result Poisson errors in the PSF at the radius of the pulsar are not neglible. However, brighter, unsaturated stars are several times further away and some change in the shape of the wings of the PSF can be discerned. We therefore used two stars as templates, the nearby Star D and the one magnitude brighter, but more distant, Star $\alpha$ (see Figure 2), and compared results.

The two template stars and their surrounding sky were independently shifted to the position of Star X using a spline interpolant program, scaled and subtracted. Although both resulting images show an increase in noise where the core of Star X had been, the mean and standard deviation of the pixels in a 10 × 10 box around the position of the pulsar agree with that expected from the undisturbed sky background.

While the presence of light from the pulsar binary is not immediately evident from a visual examination of the subtracted image at $\phi = 0.27$, the information provided by the $\phi = 0.44$ image allows us to do a sensitive statistical test for emission. The positions of stars C, D and X on the image taken at minimum agree with those on the phase $\phi = 0.44$ image to better than than 0.1 pixel; furthermore, no change in the PSF is evident between the two observations. We can therefore use the PSF of the pulsar system in the phase $\phi = 0.44$ image as a matched filter with which to convolve the two subtracted images. Specifically, we took the central 3x3 pixels of the pulsar in the $\phi = 0.44$ observations, and convolved this kernal with the data in the $\phi = 0.27$ image. This procedure results in a nearly optimal weighting for any faint underlying point source. (We did not use a shifted image of a bright star as the matched filter because the reduction in poisson noise would be far outweighed by our inability to accurately shift the underresolved core of the PSF.) After convolution, we find the central pixel at the position of the pulsar to be $3.3\sigma$ above the mean sky background in the image from which Star D was used as a template to subtract star X, and $2.4\sigma$ above the mean in the image from which Star $\alpha$ was used as a template. As a further check on these statistics, we extracted from each image the pixels in a $100 \times 100$ box which included the pulsar but excluded the background star. In both cases, we find that less than one in several hundred pixels from the extracted region is as bright as the pixel at the pulsar position, and that pulsar is at a local maximum. It is highly likely, then, that we have detected emission from the binary near optical minimum.

Due to the extended wings of the Space Telescope's PSF, it is impracticable to perform photometry on faint objects using an aperature containing most of the star's light. Instead, one must measure the light falling in a central region of the PSF, and extrapolate to the total light using the PSF's of nearby, bright unsaturated stars. Applying this method to the central $3 \times 3$ pixels of the PSF we find the flux density from the companion at minimum to be $2.1 \pm 0.6 \times 10^{-29}$ ergs s$^{-1}$ cm$^{-2}$ Hz$^{-1}$ at an effective wavelength of $7900\text{Å}$, or, in the Gunn color system, $i = 23.4 \pm 0.3$. As a check on our photometry, we can use the same extrapolation to estimate the magnitude of Star X, which has been repeatedly observed from earth at pulsar optical minimum. We find agreement with ground-based photometry (Fruchter *et al.* 1988) to a few hundredths of a magnitude.

## 3. Discussion

Unfortunately, these observations do not yet allow us to determine whether, at minimum, we are observing optical emission from a warm posterior due to heat flux from the interior of the star, or rather a crescent of the companion heated by the pulsar and visible due to the inclination on the sky of the binary's orbit. Nonetheless, we can use these observations to set an upper limit on the temperature of the side of the companion facing away from the pulsar.

As discussed in the Introduction, ground-based observations of the companion at maximum (Fruchter *et al.* 1988; van Paradijs *et al.* 1988; Djorgovski & Evans 1988; Callanan, Charles & van Paradijs 1989; Callanan, van Paridijs & Regelink 1994) provide both magnitude and color information. As a result, if we are willing to assume both a distance and extinction to the system, the size of the companion can be estimated. Fortunately the dispersion measure to the pulsar (Kulkarni, Djorgovski & Fruchter 1988), optical spectroscopy of the bow shock in the interstellar medium surrounding the pulsar (Aldcroft, Romani & Cordes 1992) and arguments based on the companion's temperature and pulsar irradiation (Djorgovski & Evans 1988) all suggest that the extinction is low, with $A_V \leq 1$. Therefore, to a good approximation, the estimated size of the companion scales approximately linearly with the assumed distance to the pulsar (Djorgovski & Evans 1988). The expected apparent magnitude of the companion at a given orbital phase is then independent of the assumed distance to the system and is only a function of the temperature across the visible face of the companion and the stellar atmospheric model.

Assuming a distance to the pulsar of about 1.2 kpc (Aldcroft, Romani & Cordes 1992; Fruchter & Goss 1992), and a companion of solar metallicity, and employing the color/temperature grid from the atmospheric model of Allard and Hauschildt (1995) we find that the flux density of our $\sim 3\sigma$ detection corresponds to a photospheric temperature of $2800 \pm 150$ K. The formal error in this estimate is dominated by the photometric and distance uncertainties. Altering the assumed metallicity by a factor of ten only changes the temperature estimate by about $1\sigma$. We caution the reader, however, that low temperature atmospheric models have so far received only limited observational testing.

If part of the light observed at minimum is contributed by a hot crescent, or if our $\sim 3\sigma$ detection is an unfortunate statistical fluke, then the coldest point on the surface of the companion must, of course, be cooler still. However, the temperature suggested by our assumptions is particularly intriguing because the implied intrinsic luminosity, $L \sim 6 \times 10^{-3}$ $L_\odot$, is approximately equal to the maximum intrinsic luminosity allowed by the best fit photometric model of Callanan *et al.* (1994) and is comparable to, but somewhat greater than, the intrinsic luminosity predicted by the Applegate and Shaham (1994) tidal dissipation model. Both of these models require that the companion be bloated beyond the radius of a degenerate star and nearly fill its Roche lobe.

It should be possible to improve upon these observational results using the repaired HST. If our detection is real, new observations would produce high signal-to-noise images in several colors, allowing a determination of the magnitude and color of the companion at conjunction. This would remove the major limitation of the analysis of Callanan *et al.* (1994) — the lack of magnitude and color information near minimum — and should permit an accurate estimate of the size of the companion, the inclination of its orbit, and its intrinsic luminosity.

– 7 –

## 4. Acknowledgements

This work was supported by a General Observer grant from NASA awarded through the Space Telescope Science Institute. ASF received additional support from a Hubble Fellowship also awarded by NASA through STScI. CDB is partially supported by NASA grant NAGW-2469, and a National Young Investigator grant from the NSF.

Figure Captions

Figure 1: Three images of the PSR B1957+20 system taken by HST at different orbital phases. Figure 1a, taken at superior conjunction in the F569W or "V" filter, shows the sytem at maximum. Figure 1b displays the system shortly before pulsar descending node, or quadrature, in the F791W or "I" filter. The background star to the north east (upper left) has the same colors as the pulsar companion at maximum, and thus provides a reference for the eye. It is this star which has confounded ground-based observations near minimum. Figure 1c was taken at inferior conjunction, or minimum in the F791W filter.

Figure 2: The field of the Planetary Camera. Stars X, C and D can be seen in Figure 1. Stars D and $\alpha$ were used as PSF templates for subtraction of the contribution of Star X at the position of PSR 1957+20.

TABLE 1
PC Observations of PSR B1957+20

| Date (UT) | Orbital Phase | Filter | Integration Time (s) |
|---|---|---|---|
| 19 OCT 91 | 0.741 – 0.759 | F569W | 600 |
| 20 OCT 91 | 0.250 – 0.265 | F791W | 500 |
| 20 OCT 91 | 0.277 – 0.292 | F791W | 500 |
| 20 OCT 91 | 0.424 – 0.440 | F791W | 500 |
| 20 OCT 91 | 0.452 – 0.467 | F791W | 500 |